\documentclass[aip,jap,preprint,showpacs]{revtex4-1}
\usepackage{graphicx}
\usepackage{graphics}

\begin{document}
	
	\title{  Study of magnetization reversal in N\'{e}el and Bloch regime of nickel and permalloy stripes using Kerr microscopy}
	
\author{Ekta Bhatia}
\email[Ekta Bhatia: ]{bhatiaekta@niser.ac.in}
\affiliation{School of Physical Sciences, National Institute of Science Education and Research (NISER) Bhubaneswar, Odisha, India-752050}
\author{Zaineb Hussain}
\affiliation{UGC-DAE Consortium for Scientific Research, Indore Centre, 452017, India}
\author{V. Raghavendra Reddy}
\affiliation{UGC-DAE Consortium for Scientific Research, Indore Centre, 452017, India}
\author{Kartik Senapati}
\affiliation{School of Physical Sciences, National Institute of Science Education and Research (NISER) Bhubaneswar, Odisha, India-752050}
	\date{\today}
	
\begin{abstract}
 We present a systematic study of the magnetization reversal of nickel and permalloy micro-stripes with N\'{e}el and Bloch domain walls using Kerr microscopy. Magnetic field driven domain propagation was observed from higher width to lower width stripes for magnetic fields applied along the length of micro-stripes. Stripe like domains were observed with nucleation starting in lower width region followed by their propagation to higher width regions for magnetic fields applied along the width of micro-stripes. The comparison of magnetization reversal in Bloch and N\'{e}el domain wall regime showed higher domain wall density in Bloch regime for both nickel and permalloy stripes.

\end{abstract}

\maketitle
\section{Introduction}

Patterned magnetic thin films is a vast area of research from several perspectives\cite{55,56,57}. First, they are a good candidate to study the effect of lateral confinement on the spin re-arrangements and magnetization reversal  processes\cite{1,2,3,4,5,6,7}; Second, they provide a basis to understand the new physical phenomena which are of fundamental interest and key to future applications in hard disk drives, magnetic random access memory (MRAM) and other spintronics devices\cite{8,9}. Using patterned magnetic films, we can control the local magnetic properties such as magnetization distribution and reversal by changing the shape and hence, magneto-static fields\cite{2,10}. The magnetization reversal of one or more ferromagnetic components is the key functional aspect of spintronics devices. For example, lateral spin valves are one of the most widely utilized device structures in spintronics\cite{14,15}. These devices involve the use of magnetic electrodes patterned in the lateral direction\cite{37}. Therefore, a systematic study of magnetization reversal for patterned ferromagnetic materials such as nickel and permalloy stripes with different aspect ratios is relevant for spintronics devices. In this context, various studies\cite{38,39,40,41,47,48,49,50,52} have been done. These studies have considerably improved the basic understanding of magnetization processes in patterned magnetic structures. In the same context, we present a comparison study of a controlled propagation of domains in connected micro-stripes for nickel and permalloy in N\'{e}el and Bloch regime of thickness. This study is also useful for domain wall superconductivity and domain wall induced spin-polarized supercurrents in superconducting-spintronics\cite{51,54}.

In this report, we investigate magnetic field driven magnetization reversal and magnetic domain patterns in N\'{e}el and Bloch regime of nickel and permalloy stripes using Kerr microscopy. We observe a controlled propagation of domains due to shape anisotropy for magnetic fields applied along the length of stripes. We observe stripe like domains for magnetic fields applied along the width of stripes.

\section{\label{sec:level1}EXPERIMENTAL DETAILS}
Nickel and permalloy thin films with thickness of 15 nm and 90 nm were prepared at room temperature using DC magnetron sputtering in an Ar pressure of 1.5 Pa using high purity ($99.99\%$) nickel and permalloy targets on cleaned Si/SiO$_{2}$ substrates. The base pressure of the deposition system was $2\times10^{-8}$ mBar. Nickel and permalloy films were patterned using photo-lithography and Ar-ion milling techniques, as shown in Fig. 1(a). The magnetic domains were imaged by high resolution magneto-optical Kerr microscope supplied by M/s Evico Magnetics, Germany. All magneto-optical Kerr effect measurements were performed at room temperature in the longitudinal mode. The ramp rate of magnetic field was kept constant as 0.1 mT/s for all the measurements. \textit{M}(\textit{H}) loops were measured by selecting an area of interest in the stripe of a particular aspect ratio. The Kerr rotation and hence, the average intensity of the domain image in the region of interest is proportional to the magnetization of that region\cite{10,11}. Therefore, the \textit{M}(\textit{H}) loop is a characteristic of an individual stripe with a specific aspect ratio. The coercive fields for the stripes of different aspect ratios were extracted from the \textit{M}(\textit{H}) loops of the respective stripes.

\section{\label{sec:level2}RESULTS AND DISCUSSIONS}

  Figure 1(a) shows the FESEM (field emission scanning electron microscope) image of the nickel (15 nm) continuous stripes with widths varying from 5 $\mu$m to 50 $\mu$m but with a fixed length of 100 $\mu$m for each stripe. The consecutive stripes are marked as R$_{1}$, R$_{2}$, R$_{3}$, R$_{4}$, R$_{5}$ and R$_{6}$ with widths of 5 $\mu$m, 10 $\mu$m, 20 $\mu$m, 30 $\mu$m, 40 $\mu$m, and 50 $\mu$m, respectively. The aspect ratio (length/width) of the respective stripes are 20, 10, 5, 3.33, 2.5, and 2, as shown in Fig. 1(a). Since the background motivation was to study the continuous propagation of domain walls, as mentioned in the abstract, experiment had to be performed in connected stripes. Generally, the domain wall energy per unit area (the sum of anisotropy, exchange and stray field energy densities) gradually increases with increasing film thickness for N\'{e}el walls, whereas for Bloch walls, the domain wall energy density decreases with increasing film thickness. Therefore, N\'{e}el walls are energetically favorable at lower thickness while Bloch walls are favorable at thickness beyond a certain threshold value\cite{21,22,23}. It has been predicted theoretically that the crossover thickness in nickel films is $\sim$50 nm\cite{24,25} whereas it is $\sim$30 nm in case of permalloy thin films\cite{26}. Therefore, throughout the text, we will refer to nickel and permalloy films of 15 nm thickness as the N\'{e}el regime and the 90 nm thickness as the Bloch domain wall regime.

  In order to compare the magnetization reversal of stripes with different aspect ratios, \textit{M}(\textit{H}) loops along with simultaneous domain images were recorded for different stripes for an in-plane applied magnetic field along the length (x-axis) and the width (y-axis) of the stripes. Henceforth, keeping in mind that the magnetic field is always applied in the plane of the substrate, throughout the text, we will refer to these two field configurations as parallel and perpendicular field, respectively. Due to the stepped pattern of the entire stripe, the corners in each segment of the stripe (corresponding to different aspect ratio) may act as a nucleation site for domains. The present experiment does not have a real control over the domain nucleation site. However, even if we assume that nucleation in a certain segment (aspect ratio) starts as a result of propagation of a domain from an adjacent segment, the full magnetization reversal in the said segment should depend on the geometry of that segment. Therefore, even though the numerical value of the obtained coercive field may not be the exact coercive field of that segment if it was isolated, the dependence of coercive field on aspect ratio, in this case, is still a relevant quantity. Fig. 1 shows the coercive fields ($H_{c}$) of nickel and permalloy stripes as a function of their width for N\'{e}el and Bloch domain wall regimes in parallel and perpendicular configuration, respectively. In parallel configuration, we observe a decrease in ($H_{c||}$) with increase in the width of the stripes. This shows that the sample geometry has a dramatic effect on the reversal of the magnetization. This is because demagnetizing fields come into picture on reducing the dimensions in patterned structures. The demagnetization field for a single stripe is proportional to $M_{s}d/w$, where $M_{s}$ is the saturation magnetization, d is the thickness and w is the width of the stripe\cite{27,28}. The demagnetization field, $H_{d}$, reduces the internal field of a stripe to $H_{i} = H_{a} - H_{d}$, where $H_{a}$ is the applied field. Therefore, the coercive field\cite{30,31,32} of stripe of width w becomes:
  \begin{center}
  	$H_{c} =H_{0} + A/w$
  \end{center}
  where $H_{0}$ is the coercive field of a stripe with infinite width (thin film) and A is a constant parameter which depends on the thickness, finite length shape anisotropy factor and saturation magnetization of the stripe. The fitting model assumes that a small part of the stripe reverses magnetization coherently  and propagates along the stripe. Here, $H_{0}$ and $A$ are free fitting parameters. The fitting results gave the $H_{0}$ values of 65 Oe and 20 Oe for nickel (15 nm) and nickel (90 nm) films while it was 25 Oe and 2 Oe for permalloy (15 nm) and permalloy (90 nm) films. The values of $H_{0}$ show the soft magnetic nature of nickel and permalloy thin films and are consistent with the literature\cite{33,34}. The stripe with lower width (higher aspect ratio) has a larger shape anisotropy and therefore, requires a larger energy to switch, resulting into higher H$_{c}$. On the other hand, in perpendicular configuration, we observe an increase in $H_{c}$ with increase in the width of stripes. Due to the shape anisotropy, the short axis of the stripes acts like a hard axis in these patterned structures. The stripe with lower width (higher aspect ratio) has a higher tendency to act like a hard axis, resulting into lower H$_{c}$. Similar results were reported in previous literature for magnetic fields applied along the easy axis and hard axis of Co nanowires\cite{35}. We emphasize here that for a pure study of the aspect ratio dependence of the coercive field one must do experiments on isolated segments, as reported earlier\cite{35}. However, one of the background motives of this work is to study the propagation of domain walls in magnetic stripe in a controllable manner so that one can design experiments where a magnetic domain wall (due to its built-in natural magnetic inhomogeneity) may be used to trigger a functional response. Since propagation is the key issue, the experiment had to be done on a single stripe where the segments with different aspect ratio are connected to each other. The important point we would like to study here is that by successively changing the aspect ratios of the segments whether we are able to get some control over the propagation of the domain walls.
 
Figure 2(a) shows the \textit{M}(\textit{H}) loop for a rectangular region marked in yellow dashed line in Fig. 2(b) for a permalloy (15 nm) stripe of aspect ratio 3.33 (marked as R$_{4}$ in Fig. 1(a)) in parallel configuration. Fig. 2(b), (c), (d) show the Kerr microscope images of magnetic domains corresponding to points A, B, C marked in Fig. 2(a). We observe a propagation of domains from higher width stripe to the adjacent lower width stripe as shown in Fig. 2(c).  

In Fig. 2(e), we show the \textit{M}(\textit{H}) loop for a rectangular region marked in yellow dashed line in Fig. 2(f) for a permalloy (15 nm) stripe of aspect ratio 3.33 in perpendicular configuration. 
We observe that domain nucleation initiates in the lower width region, at some nucleation points as shown in Fig. 2(f). Then, domains started to grow around the nucleation points and stripe like domains were formed giving rise to multiple domain walls throughout the structure, as shown in Fig. 2(g). The formation of multiple domain walls may be due to the minimization of demagnetization energy along the width of stripe. We notice that the sequence in which the domain nucleation starts in different stripes, follows the variation of coercive fields as a function of width shown in Fig. 1(h). 
     
Figure 3(a) shows the \textit{M}(\textit{H}) loop for a rectangular region marked in yellow dashed line in Fig. 3(b) for a permalloy (15 nm) stripe of aspect ratio 10 (marked as R$_{2}$ in Fig. 1(a)) in parallel configuration.  We observe that domain nucleation initiates in the higher width (lower aspect ratio) region, at some nucleation points, as shown in Fig. 3(c). A monotonic decrease in the coercivity of alternate stripes results in a propagation of domains from one stripe to other as shown in Fig. 3(c), (d). The comparison of domain propagation in Fig. 2 and Fig. 3 shows that the propagation is more directional and hence, controlled for stripes of higher aspect ratio in parallel configuration.  

In Fig. 3(f), we show the \textit{M}(\textit{H}) loop for a rectangular region marked in yellow dashed line in Fig. 3(g) for a permalloy (15 nm) stripe of aspect ratio 10 in perpendicular configuration. 
  We observe multiple domain walls throughout the structure, as shown in Fig. 3(g), 3(h). The comparison of Kerr images in Fig. 2 and Fig. 3 shows similar stripe domain patterns for different aspect ratios in perpendicular configuration.  
 
 Figure 4(a) shows the \textit{M}(\textit{H}) loop for a rectangular region marked in yellow dashed line in Fig. 4(b) for a nickel (15 nm) stripe of aspect ratio 3.33 in parallel configuration. Note that the Kerr images shown in this manuscript have been taken from stripes of aspect ratio 3.33 for representation and comparison purpose. We observe a propagation of domains in the parallel configuration as shown in Fig. 4(c) and (d). 
 Since the growth of domains happens in a particular direction, we propose that a 180$^{\circ}$ domain wall propagates from wider stripes to narrower stripes of the patterned geometry. R.H.Wade\cite{29} in 1962 showed that the polycrystalline nickel thin film of 16 nm thickness comprises of 180$^{\circ}$ domain walls with a domain wall thickness of 55$\pm$0.5 nm. The comparison of domain propagation in permalloy (15 nm) in Fig. 2 and nickel (15 nm) in Fig. 4 shows that the propagation is more gradual in nickel compared to permalloy.
 
  In Fig. 4(f), we show the \textit{M}(\textit{H}) loop for a rectangular region marked in yellow dashed line in Fig. 4(g) for a stripe of aspect ratio 3.33 in perpendicular configuration. The magnetization reversal resulted in a growth of stripe like domains, as shown in Fig. 4(h), 4(i). The comparison of Kerr images in Fig. 2(g) and 4(h) shows that the domain patterns are similar irrespective of the material in perpendicular configuration. This might be due to the dominance of shape anisotropy in the process of domain formation.

Figure 5(a) shows the \textit{M}(\textit{H}) loop for a rectangular region marked in yellow dashed line in Fig. 5(b) for a permalloy (90 nm) stripe of aspect ratio 3.33 in parallel configuration. Similar to the case of permalloy (15 nm), domain wall propagates from the stripe with higher width (lower aspect ratio) to the stripe with lower width (higher aspect ratio), as shown in Fig. 5(c) and (d). In Fig. 5(f), we show the \textit{M}(\textit{H}) loop for a rectangular region marked in yellow dashed line in Fig. 5(g) for a stripe of aspect ratio 3.33 in perpendicular configuration.  Stripe like domains were observed in perpendicular configuration, as shown in Fig. 5(g), (h).  The comparison of Kerr images shown in Fig. 2(g) and 5(g) suggests that the domain width decreases in Bloch regime compared to N\'{e}el regime and hence, the domain wall density increases in Bloch regime. This may be because of an increase of demagnetization energy with increasing thickness, as evident from the experssion $H_{d} = M_{s}d/w$.

  Figure 6(a) shows the \textit{M}(\textit{H}) loop for a rectangular region marked in yellow dashed line in Fig. 6(b) for a stripe of aspect ratio 3.33 in parallel configuration. We observe a propagation of domains from higher width region to lower width region as shown in Fig. 6(c), (d).
In Fig. 6(f), we show the \textit{M}(\textit{H}) loop for a rectangular region marked in yellow dashed line in Fig. 6(g) for a stripe of aspect ratio 3.33 in perpendicular configuration.
 Stripe like domains were observed in perpendicular configuration, as shown in Fig. 6(g).  The comparison of Kerr images shown in Fig. 4(i) and 6(g) suggests that the domain width decreases and hence, the domain wall density increases in Bloch regime compared to N\'{e}el regime. This result is similar to the one obtained from comparison of N\'{e}el and Bloch regime in case of permalloy. The comparison of domain images in parallel and perpendicular configuration shows a higher domain wall density in perpendicular configuration in all the cases discussed above.

\section{\label{sec:level3}Conclusion}

We have investigated magnetic field driven magnetization reversal and imaged magnetic domains in nickel and permalloy stripes with N\'{e}el and Bloch domain walls in different field orientations by Kerr microscopy. We observe propagation of domain walls in parallel field configuration and growth of multiple stripe like domains in perpendicular field configuration. We observed more controlled propagation of domains in stripes with higher aspect ratio. The comparison of stripes with N\'{e}el and Bloch domain walls showed higher domain wall density in Bloch regime. From these results, we conclude that it is advantageous to use Bloch domain walls and perpendicular configuration in studies requiring higher domain wall density such as domain wall magneto-resistance, domain wall switching etc. in spintronics\cite{43,44}.  This study may find an importance in the field of domain wall superconductivity and domain wall induced triplet superconductivity\cite{51,13,42}.

\begin{acknowledgments}
We acknowledge the funding from National Institute of Science Education and Research(NISER), DST-Nanomission (SR/NM/NS-1183/2013) and DST SERB (EMR/2016/005518) of Govt. of India. 

\end{acknowledgments}

  \begin{figure}[ht]
	\centering
	\includegraphics[width=14cm]{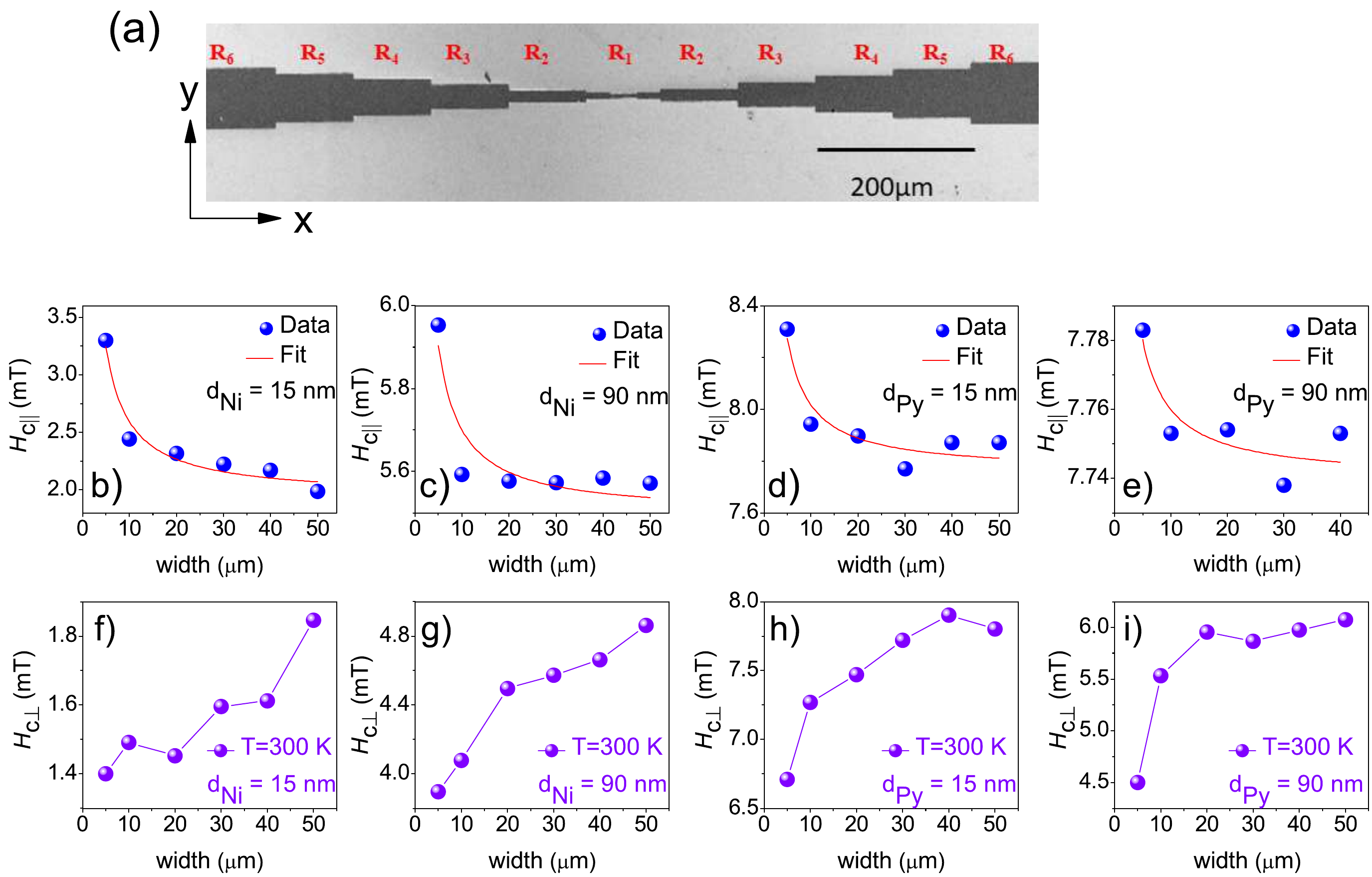}\\
	%
	\caption{a) FESEM image of the nickel stripes with thickness of 15 nm, length of 100 $\mu$m each and widths varying from 5 $\mu$m to 50 $\mu$m; Variation of coercive fields with respect to width of nickel and permalloy stripes have been shown along with the fitting in (b) - (e), coercive fields have been taken from the hysteresis loops recorded using MOKE measurements of these stripes. $H_{c||}$ represents the coercive field for magnetic fields applied along the length (x-axis) of stripes of nickel films with thickness of b) 15 nm, c) 90 nm and for permalloy films with thickness of d) 15 nm, e) 90 nm; $H_{c\perp}$ represents the coercive field for magnetic field applied along the width (y-axis) of the stripes for nickel films with thickness of f) 15 nm, g) 90 nm and for permalloy films with thickness of h) 15 nm, i) 90 nm}\label{Fig.2}
\end{figure}  
 \begin{figure}[ht]
	\centering
	\includegraphics[width=18cm]{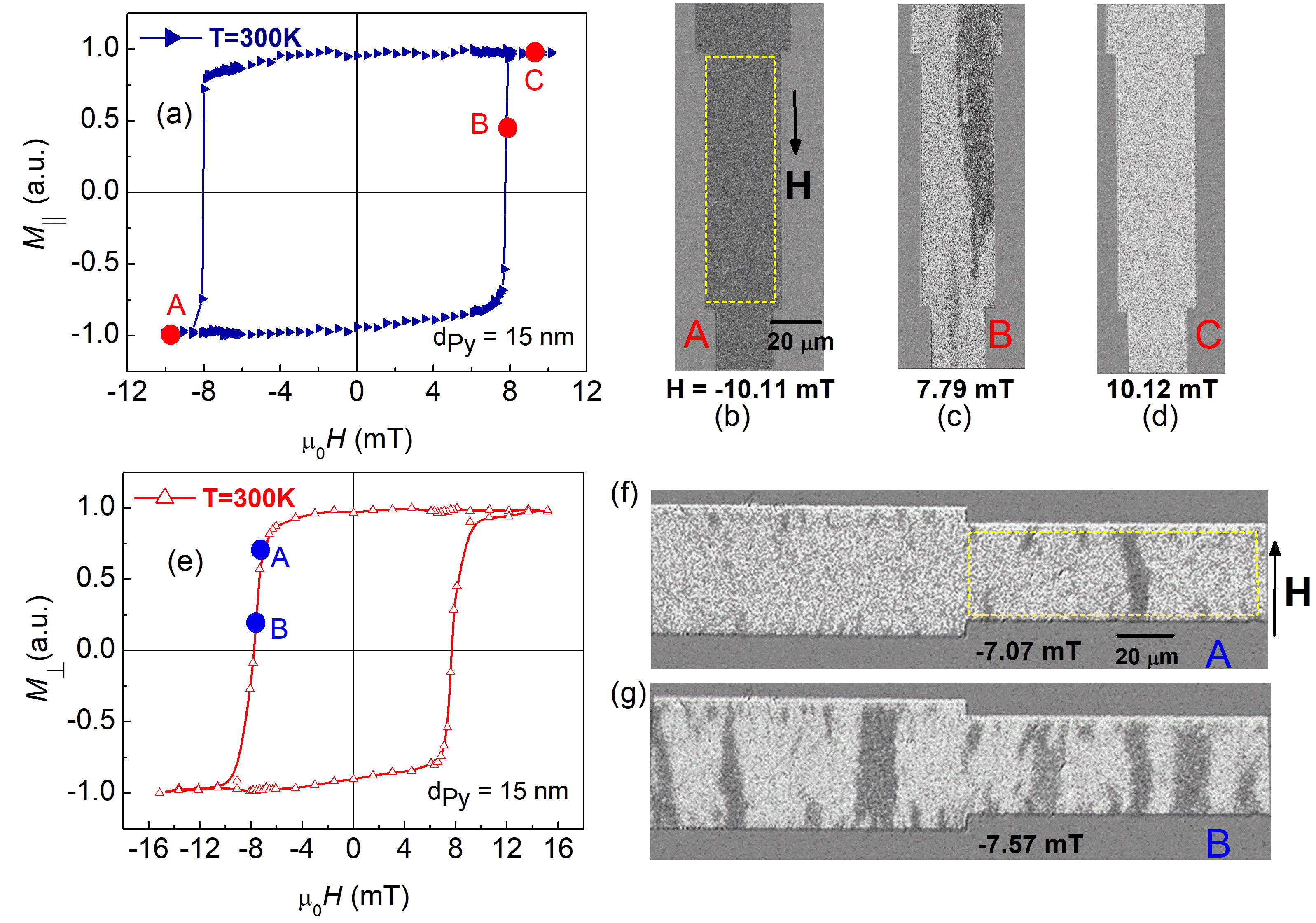}\\
	%
	\caption{ (a) \textit{M}(\textit{H}) loop for a rectangular region marked in yellow dashed line in (b) for a permalloy stripe of thickness 15 nm and aspect ratio 3.33 in parallel configuration; (b) to (d) are the Kerr	images corresponding to the field points marked A-C in (a), respectively, (e) \textit{M}(\textit{H}) loop for a rectangular region marked in yellow dashed line in (f) in a permalloy stripe of thickness 15 nm and aspect ratio 3.33 for perpendicular configuration; (f) and (g) are the Kerr images corresponding
		to the field points marked A, B in (e); Black arrows in (b) and (f) represent the direction of
		applied magnetic field (H). The scale bar shown in (b) and (f) are valid for all the images. The bright and dark portions of the image represent domains oriented in the opposite direction.
}\label{Fig.7}
\end{figure}

\begin{figure}[ht]
	\centering
	\includegraphics[width=14cm]{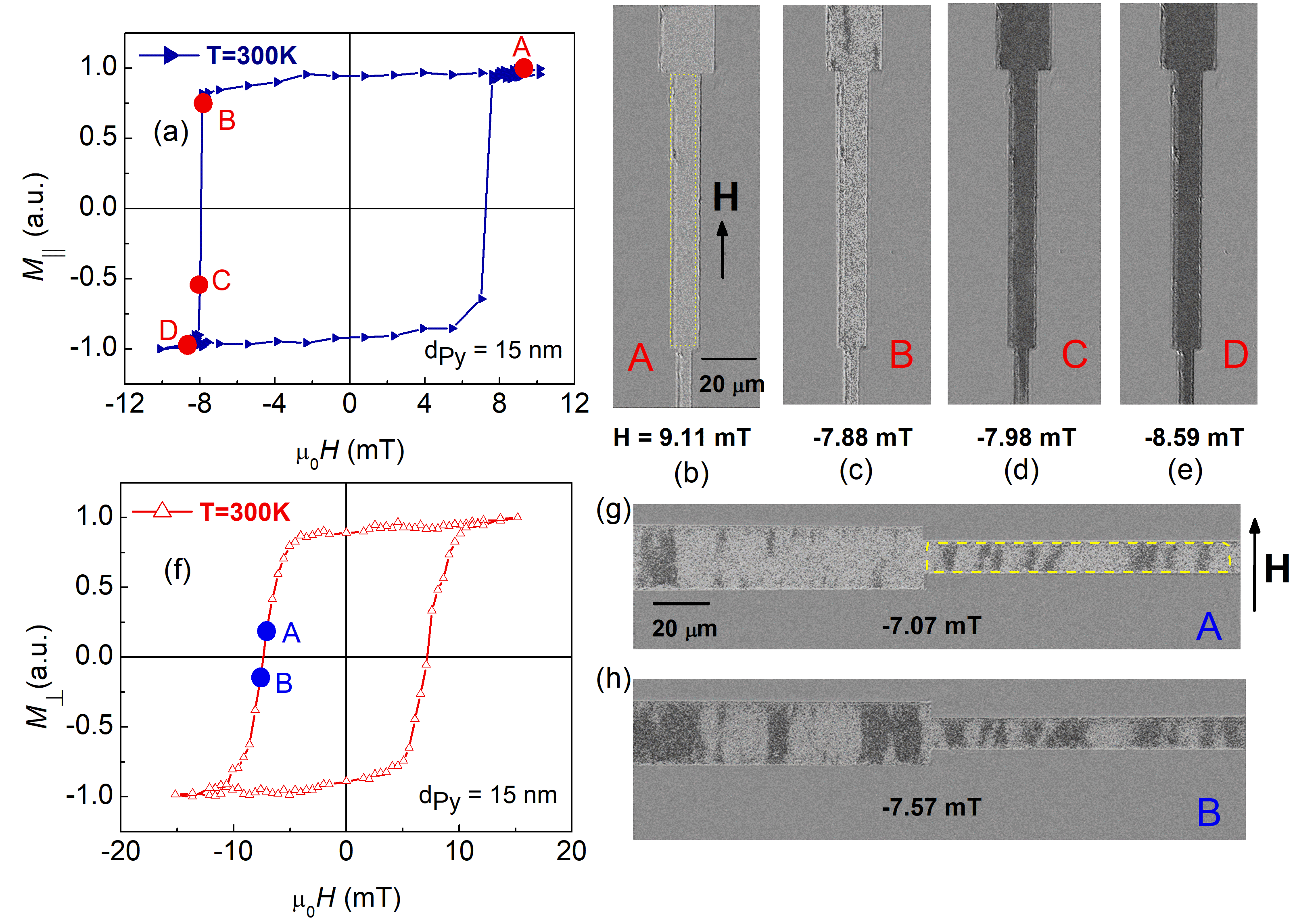}\\
	%
	\caption{\textit{M}(\textit{H}) loop for a rectangular region marked in yellow dashed line in (b) for a permalloy stripe of thickness 15 nm and aspect ratio 10 in parallel configuration;  (b) - (e) are the Kerr images corresponding to the field points marked A - D in (a), respectively, (f) \textit{M}(\textit{H}) loop for a rectangular region marked in yellow dashed line in (g) for a permalloy stripe of thickness 15 nm and aspect ratio 10 in perpendicular configuration;  (g) and (h) are the Kerr images corresponding to the field points marked A, B in (f). The scale bar shown in (b) and (g) are valid for all the images.}\label{Fig.9}
\end{figure}
 
\begin{figure}[ht]
	\centering
	\includegraphics[width=14cm]{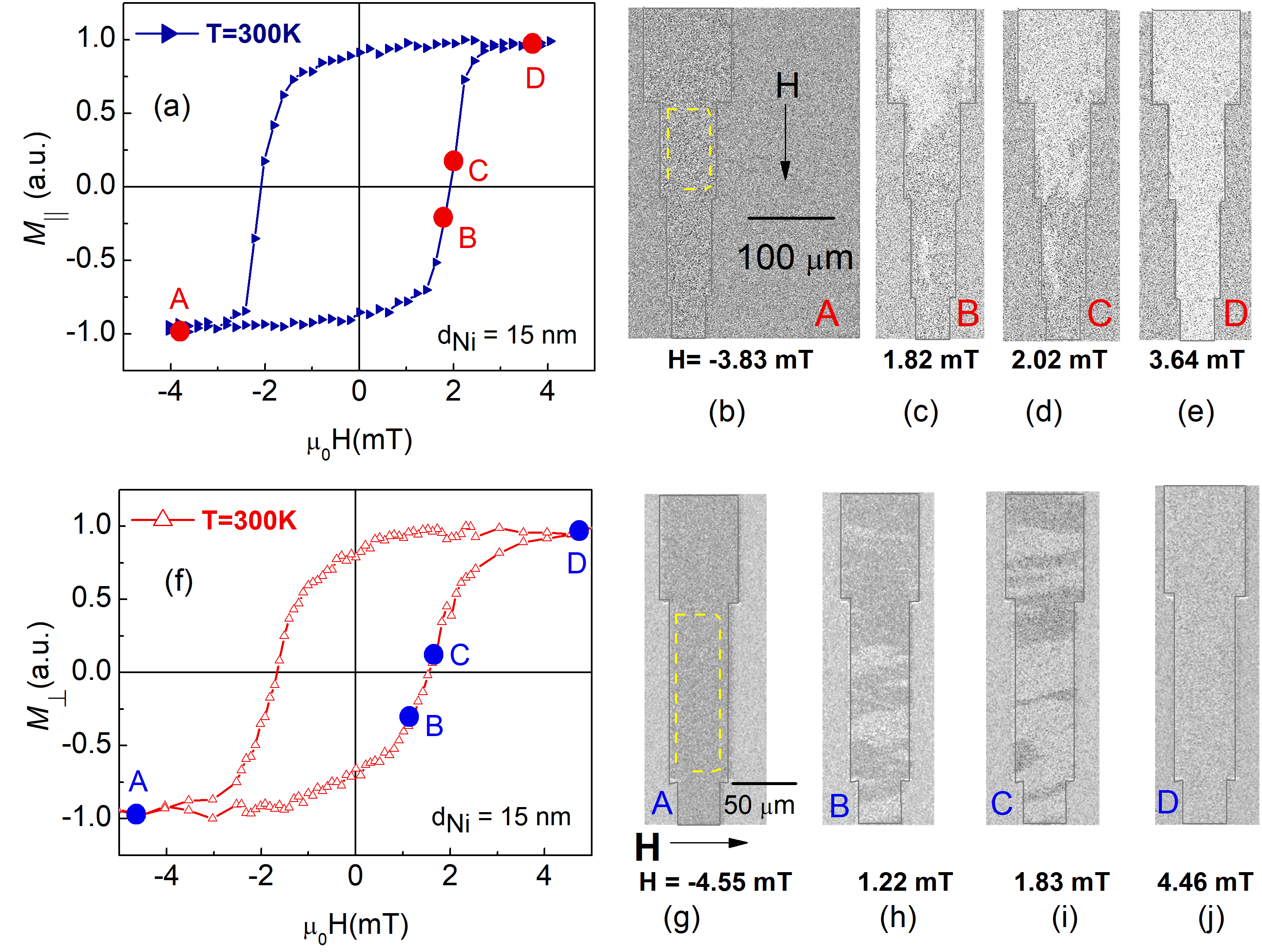}\\
	%
	\caption{\textit{M}(\textit{H}) loop for a rectangular region marked in yellow dashed line in (b) for a nickel stripe of thickness 15 nm and aspect ratio 3.33 in parallel configuration;  (b) - (e) are the Kerr images corresponding to the field points marked A - D in (a), respectively, (f) \textit{M}(\textit{H}) loop for a rectangular region marked in yellow dashed line in (g) in a nickel stripe of thickness 15 nm and aspect ratio 3.33 in perpendicular configuration;  (g) - (j) are the Kerr images corresponding to the field points marked A - D in (f). The scale bar shown in (b) is valid for images (b) - (e). The scale bar shown in (g) is valid for images (g) - (j).}\label{Fig.9}
\end{figure}
\begin{figure}[ht]
	\centering
	\includegraphics[width=14cm]{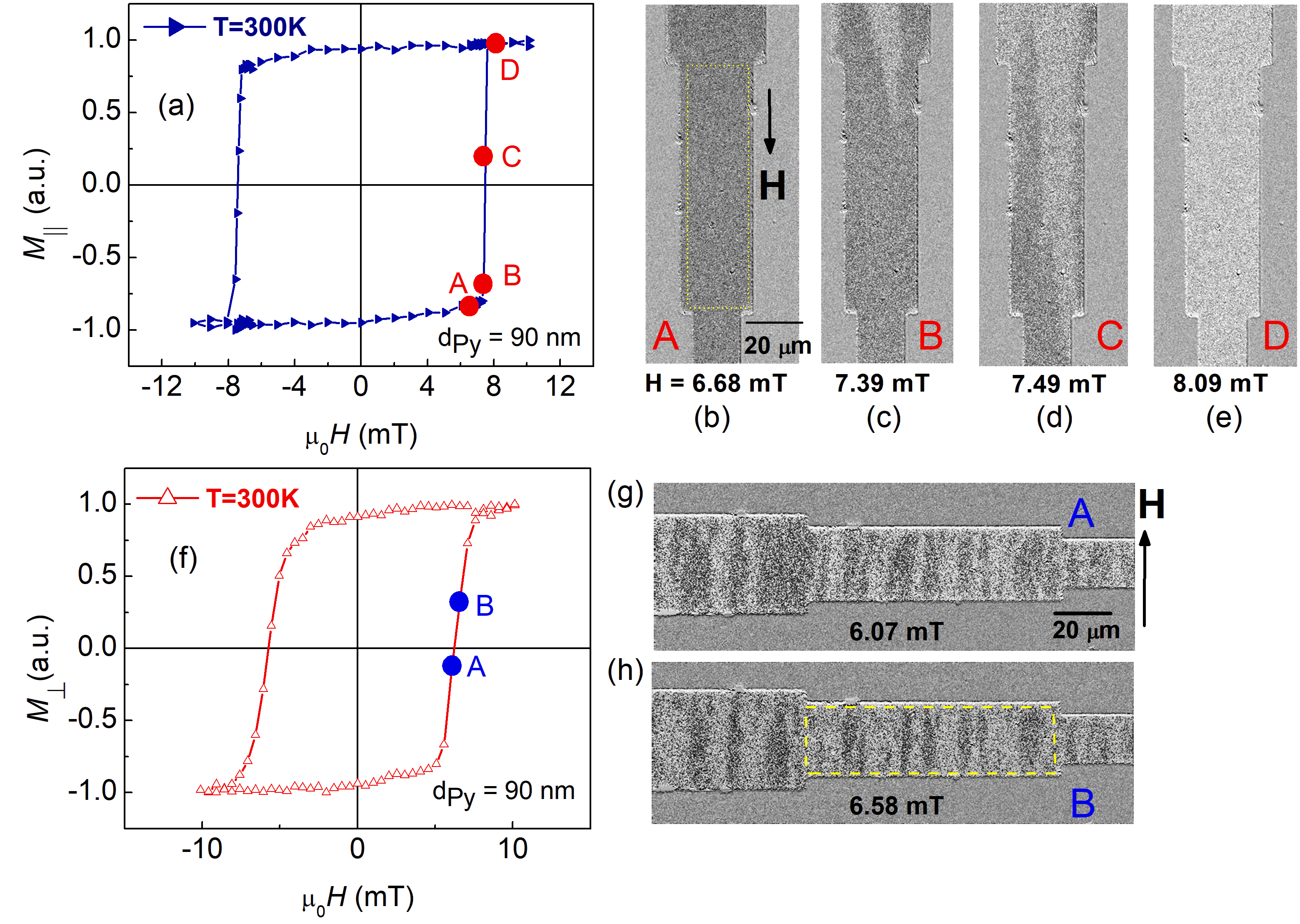}\\
	%
	\caption{(a) \textit{M}(\textit{H}) loop for a rectangular region marked in yellow dashed line in (b) for a permalloy stripe of thickness 90 nm and aspect ratio 3.33 in parallel configuration; (b) - (e) are the Kerr images corresponding to the field points marked A - D in (a), respectively, (f) \textit{M}(\textit{H}) loop for a rectangular region marked in yellow dashed line in (g) in a permalloy stripe of thickness 90 nm and aspect ratio 3.33 in perpendicular configuration; (g), (h) represent the Kerr images corresponding to the field point marked A, B in (f). The scale bar shown in (b) and (g) are valid for all the images.}\label{Fig.8}
\end{figure}
 \begin{figure}[ht]
	\centering
	\includegraphics[width=14cm]{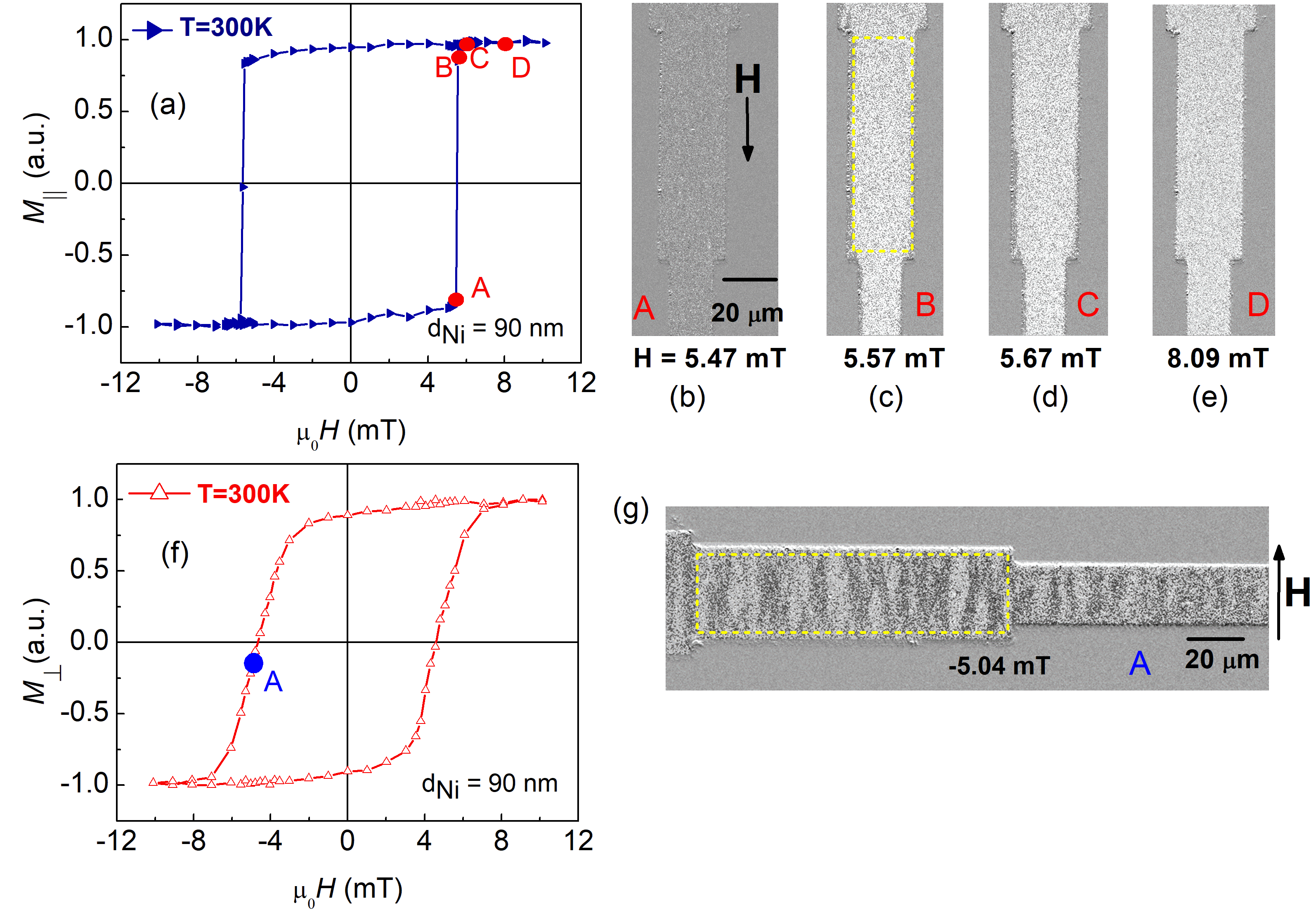}\\
	%
	\caption{\textit{M}(\textit{H}) loop for a rectangular region marked in yellow dashed line in (b) for a nickel stripe of thickness 90 nm and aspect ratio 3.33 in parallel configuration;  (b) - (e) are the Kerr images corresponding to the field points marked A - D in (a), respectively, (f) \textit{M}(\textit{H}) loop for a rectangular region marked in yellow dashed line in (g) in a nickel stripe of thickness 90 nm and aspect ratio 3.33 in perpendicular configuration;  (g) represents the Kerr image corresponding to the field points marked A in (f). The scale bar shown in (b) and (g) are valid for all the images.}\label{Fig.10}
\end{figure}

\begin{thebibliography}{1}\footnotesize
\bibitem{55} C. L. Dennis, R. P. Borges, L. D. Buda, U. Ebels, J. F. Gregg, M. Hehn, E. Jouguelet, K.
	Ounadjela, I. Petej, I. L. Prejbeanu, and M. J. Thornton, J. Phys.: Condens. Matter 14 (2002), R1175 .
\bibitem{56} O. Hellwig, L. J. Heyderman, O. Petracic, and H. Zabel,	“Competing Interactions in Patterned and Self-Assembled Magnetic Nanostructures,” in Magnetic Nanostructures, H. Zabel and M. Farle, eds., Springer Berlin Heidelberg, (2013) Vol. 246, p. 189-234.
\bibitem{57} M. Kl\"{a}ui, J. Phys. Cond. Mat. 20 (2008), p. 313001.
\bibitem{1} R. P. Cowburn, D. K. Koltsov, A. O. Adeyeye, M. E. Welland, D. M. Tricker, Phys. Rev. Lett. 83 (1999), p. 1042.
\bibitem{2} R. P. Cowburn, J. Phys. D: Appl. Phys. 33 (2000) R1-R16.
\bibitem{3}  M. Schneider, H. Hoffmann, S. Otto, Thu. Haug, J. Zweck, J. Appl. Phys. 92 (2002), p. 1466.
\bibitem{4} J. K. Ha, R. Hertel, J. Kirschner, Phys. Rev. B 67 (2003), p. 224432.
\bibitem{5} A. Wachowiak, J. Wiebe, M. Bode, O. Pietzsch, M. Morgenstern, R. Wiesendanger, Science 298 (2002), p. 577.
\bibitem{6} M. E. Schabes, H. N. Bertram, J. Appl. Phys. 64 (1988), p. 1347.
\bibitem{7} J. K. Ha, R. Hertel and J. Kirschner, Phys. Rev. B 67 (2003), p. 064418.
\bibitem{8} J. Ferre, Dynamics of Magnetization Reversal: From Continuous to Patterned Ferromagnetic
Films, edited by Hillebrands B. and Ounadjela K. (Springer) 2002, p. 127.
\bibitem{9} R. Victora, The Physics of Ultra-High-Density Magnetic Recording, edited by Plumer
M. L., van Ek J. and Weller D., Perpendicular Recording Media (Springer) (2001), p. 230.
\bibitem{10} A. Hubert, R. Schafer, Magnetic Domains, Springer, Berlin, (2008).
\bibitem{14} H. C. Koo, J. H. Kwon, J. Eom, J. Chang, S. H. Han, and M. Johnson, Science 325 (2009), p. 1515.
\bibitem{15} F. J. Jedema, H. B. Heersche, A. T. Filip, J. J. A. Baselmans, B. J. van Wees, Nature London 416 (2002), p. 713.
\bibitem{37} Frank Br\"{u}ssing, Gregor Nowak, Alexandra Schumann, Stefan Buschhorn, Hartmut Zabel and Katharina Theis-Br\"{o}hl, J. Phys. D: Appl. Phys. 42 (2009) p. 165001.
\bibitem{38} Thomas Schrefl, Josef Fidler, K.J. Kirk, J.N. Chapman, J. Magn. Magn. Mater. 175 (1997), p. 193.
\bibitem{39} J. Ferr\'{e}, in Spin Dynamics in Confined Magnetic Structures I edited by B. Hillebrands and K. Ounadjela (Springer-Verlag, Berlin, Heidelberg, 2002), Vol. 83, p. 127–168.
\bibitem{40} R. H. Koch, J. G. Deak, D. W. Abraham, P. L. Trouilloud, R. A. Altman, Yu Lu, W. J. Gallagher, R. E. Schenerlein, K. P. Roche, S.S.P Parkin, Phys. Rev. Lett. 81 (1998), p. 4512.
\bibitem{41} A. Fern\'{a}ndez-Pacheco, J. M. De Teresa, A. Szkudlarek, R. C\'{o}rdoba, M. R. Ibarra, D. Petit, L. O$'$Brien, H. T. Zeng, E. R. Lewis, D. E. Read and R. P. Cowburn, Nanotechnology 20 (2009), p. 475704.
\bibitem{47} C.L. Dennis, R. P. Borges, L. D. Buda, U. Ebels, J. F. Gregg, M. Hehn, E. Jouguelet, K. Ounadjela, I. Petej, I. L. Prejbeanu and M. J. Thornton, J. Phys.: Condens. Matter 14 (2002), p. R1175-R1262.
\bibitem{48} O. Hellwig et al. Springer Tracts in Modern Physics, Magnetic Nanostructures, (2013) Vol. 246, p. 189-234;
\bibitem{49} M. Kl\"{a}ui, J. Phys. Cond. Mat. 20 (2008), p. 313001.
\bibitem{50} Andr\'{e} Bisig, M. St\"{a}rk, M.-A. Mawass, C. Moutafis, J. Rhensius, J. Heidler, F. B\"{u}ttner, M. Noske, M. Weigand, S. Eisebitt, T. Tyliszczak, B. V. Waeyenberge, H. Stoll, G. Sch\"{u}tz and M. Kl\"{a}ui, Nat. Commun. 4  (2013), p. 2328.
\bibitem{52} J. McCord, J. Phys. D: Appl. Phys. 48 (2015), p. 333001.
\bibitem{51} J. W. A. Robinson, F. Chiodi, M. Egilmez, G\'{a}bor B. Hal\'{a}sz and M. G. Blamire  Sci. Rep. 2 (2012), p. 699.
\bibitem{54}  E. Bhatia, Z. H. Barber, I. J. Maasilta, and K. Senapati, AIP Advances 9 (2019), p. 045107.
\bibitem{11} R. Sch\"{a}fer, $"$Investigation of domains and dynamics of domain walls by the magneto-optical Kerr-effect$"$, in Handbook of Magnetism and Advanced Magnetic Materials, Helmut Kronmuller, Stuart Parkin (Editor), John Wiley \& Sons, (2007).
	\bibitem{21}  L. N$\acute{e}$el, C.R. Acad. Sci. (Paris) 241 (1955), p. 533.
	\bibitem{22} F. Bloch, Z. Phys. 74 (1932), p. 295.
	\bibitem{23} S. Middelhoek, J. Appl. Phys. 34 (1963), p. 1054.
	\bibitem{24}  J. Silcox, Phil. Mag. 8 (1963), p. 7. 
	\bibitem{25} C. T. Hsieh, J. Q. Liu, J. T. Lue, Appl. Surf. Sci. 252 (2005), p. 1899.
\bibitem{26} T. Trunk and M. Redjdal, A. K\'{a}kay, M. F. Ruane, F. B. Humphrey, J. Appl. Phys. 89 (2001), p. 7606.
\bibitem{27} B. Hausmanns, T. P. Krome, G. Dumpich, E. F. Wassermann, D. Hinzke, U. Nowak and K. D. Usadel, J. Magn. Magn. Mater. 240 (2002), p. 297.
\bibitem{28} W. C. Uhlig and J. Shi, Appl. Phys. Lett. 84 (2004), p. 759.

\bibitem{30} S. W. Yuan, H. N. Bertram, J. F. Smyth, S. Schultz, IEEE Trans. Magn. 28 (1992), p. 3171-3173.
\bibitem{31} L.A. Rodr\'{\i}guez, L. Deen, R. C\'{o}rdoba, C. Mag\'{e}n, E. Snoeck, B. Koopmans, J.M. De Teresa, Beilstein J. Nanotechnol. 6 (2015), p. 1319-1331.
\bibitem{33} W. Rattanasakulthong, P. Sirisangsawang, S. Pinitsoontorn, and C. Sirisathitkul, Adv. Mat. Res 335-336 (2011), p. 1443–1447.
\bibitem{34} I. W. Wolf, J. Electrochem. Soc. 108 (1961), p. 959. 
		\bibitem{35} G. Kartopu, O. Yal\c{c}ın, M. Es-Souni, and A. C. Ba\c{s}aran, J. Appl. Phys. 103 (2008), p. 093915 .
	\bibitem{29} R.H. Wade, Proc. Phys. Soc. 79 (1962), p. 1237. 
\bibitem{43} G. X. Miao, M. D. Mascaro, C. H. Nam, C. A. Ross, and J. S. Moodera, Appl. Phys. Lett. 99 (2011), p. 032501.
\bibitem{44}  J. Tr\"{u}tzschler, K. Sentosun, B. Mozooni, R. Mattheis, and J. McCord, Sci. Rep. 6 (2016) p. 30761.
\bibitem{13} Jacob Linder, Klaus Halterman, Phys. Rev. B 90, (2014), p. 104502.
\bibitem{42} Z.R. Yang, M. Lange, A. Volodin, R. Szymczak, V.V. Moshchalkov, Nat. Mater., 3 (2004), p. 793.	
\end{thebibliography}
\end{document}